\documentclass{aa}
\usepackage{graphicx,natbib}
\usepackage{epsfig} 
\newcommand{\lya}{Ly$\alpha$}
\newcommand{\etal}{et al.}
\newcommand{\ha}{H$\alpha$}
\newcommand{\ecs}{erg cm$^{-2}$ s$^{-1}$}

\begin{document} 
\title{A Chandra study of X-ray sources in the field of the z=2.16 radio
galaxy MRC 1138-262}

\author{L. Pentericci\inst{1} \and J.D. Kurk\inst{2} \and
C.L.Carilli\inst{3} \and D.E. Harris\inst{4} \and G.K. Miley\inst{2}
\and H.J.A. R\"ottgering\inst{2}}

\institute{
             Max-Planck-Institut f\"ur Astronomie, K\"onigstuhl 17,
             D-69117, Heidelberg, Germany
\and 
             Sterrewacht Leiden, P.O. Box 9513, 2300 RA, Leiden, 
             The Netherlands               
\and
             National Radio Astronomy Observatory, P.O. Box 0, 
             Socorro NM, 87801, USA
\and 
             Smithsonian Astronomical Observatory, Harvard-Smithsonian
             Center for Astrophysics, 60 Garden Street, Cambridge, MA 02138
}

\offprints{L.Pentericci, {laura@mpia.de}}

\abstract{ 

We present results from a {\it Chandra X-ray Observatory} study of the
field X-ray source population in the vicinity of the radio galaxy MRC
1138-262.  Many serendipitous X-ray sources are detected in an area of
8$'\times$8$'$ around the radio source and 90\% are identified in our
deep VLT images.  The space density of such sources is 
higher than expected on the basis of the statistics of {\it ROSAT} and
{\it Chandra} deep surveys.  The most likely explanation is in terms
of a concentration of AGN associated with the protocluster at $z =
2.16$ which was found around the radio galaxy in previous studies.
Two sources have a confirmed spectroscopic redshift close to that of
the radio galaxy, and for three more sources other observations
suggest that they are associated with the protocluster.  
Four of these
five X-ray sources form, together with the radio galaxy, a filament in
the plane of the sky. The direction of the filament is similar to that
of the radio source axis, the large scale distribution of the other
protocluster members, the 150 kpc-sized emission-line halo and the extended X-ray emission associated with
the radio galaxy.

The majority of optically identified X-ray sources in this field have
properties consistent with type I AGN, a few could be soft, low
luminosity galaxies, one is probably an obscured (type II) AGN and one
is a star.  These statistics are consistent with the results of deep
X-ray surveys.  
}

\date{Received date / Accepted date} 

\maketitle

\section{Introduction}
Since its launch, the {\it Chandra X-ray Observatory} has been
providing the deepest and sharpest images of the X-ray sky ever: one
of its most remarkable characteristics is its high spatial resolution
($<$ 1$''$) and astrometric precision ($\sim 1''$), well matched to
typical optical and near infrared (NIR) imaging resolutions. This allows
unambiguous identifications of faint X-ray sources and the possibility
to study the morphologies of the X-ray emitting host galaxies.

We have recently observed the radio galaxy MRC 1138-262, and the
surrounding field, with {\it Chandra} to study the extended X-ray
emission associated with this radio source (Carilli et al.\ 2002),
which was previously detected with {\it ROSAT}.  This radio galaxy at
redshift 2.16 has a quite complex optical and NIR morphology,
resembling a massive galaxy in the early stages of formation and it is
embedded in a giant Ly$\alpha$ halo (Pentericci \etal\ 1997).  Deep
VLT observations have shown that MRC 1138-262 resides at the center of
a protocluster consisting of at least 20 confirmed cluster members
(Pentericci \etal\ 2000, Kurk \etal\ 2002b).

{\it Chandra}'s great sensitivity allows the detection of many
X-ray emitters in the field besides the main target, 
even in observations with modest exposure times (a few
ten thousand seconds in our case).  In this paper we report on these
serendipitous X-ray sources: we find an excess of soft X-ray sources
in the field, as compared to the predictions for a non-cluster field
based on the $\log N - \log S$ relation derived from deep {\it ROSAT}
and {\it Chandra} measurements.  We present optical identifications of
the X-ray sources, discuss their nature and their possible relation to
the protocluster structure at redshift 2.16.  

Throughout the paper we assume a flat, $\Lambda$-dominated universe
with $H_{0} = 65$ km s$^{-1}$ Mpc$^{-1}$, $\Omega_M = 0.3$ and
$\Omega_{\Lambda} = 0.7$.

\section{X-ray observations} 

\subsection{Source detection}
The field of the radio galaxy MRC 1138-262 was observed for 39.5 ksec
on June 6, 2000, with the back-illuminated ACIS-S CCD detector on the
{\it Chandra X-ray Observatory}. Standard ACIS settings were used: TE
mode with 3.2 s readout and 'faint' telemetry format. The principal
aim of this observation was to study the morphology of the extended
X-ray emission associated with the central radio galaxy.  We refer to
Carilli \etal\ (2002) for a detailed description of the observations,
data processing and results on the main target.

The radio galaxy was positioned $\sim$1.5\arcmin\ from the standard aim point
on the ACIS-S3 chip.  Besides MRC 1138-262, many fainter X-ray
emitters are registered on this chip.  To localize these serendipitous
sources we applied the WAVDETECT algorithm (Freeman \etal\ 2002) of
the Chandra Interactive Analysis of Observations (CIAO) software.
Given the sensitivity of the observation, we expect to detect mostly
sources with soft X-ray spectra. For this reason and to inprove the S/N (the 
background is lower at lower energies), we have limited the energy
range for source detection to a maximum of 2 keV.  
We detect a total of 21 sources on the S3 chip, including the
nucleus of the radio galaxy (but excluding the extended emission
around this source), of which 18 have S/N $> 3$.  All sources are
clearly visible by eye.  As a check of the validity of the detections,
we have also applied the CELLDETECT algorithm of CIAO with a S/N limit
of 2.5. This algorithm recovers most of the sources, with the
exception of the faintest.  Further confirmation of the reality of the
X-ray sources comes from the fact that they all have an optical
identification.

\begin{table*}
\begin{center}
\caption{X-ray point sources in the field of MRC 1138-262 } \label{xray}
\begin{tabular}{r c c c r@{$\pm$}l c r@{$\pm$}l c c c} \hline \hline
N  & RA & Dec & $\theta$ & \multicolumn{2}{c}{Soft cnts} & Soft flux & 
\multicolumn{2}{c}{Hard cnts} & Hard flux & Aperture & HR \\
(1) & (2) & (3) & (4) & \multicolumn{2}{c}{(5)} & (6) & \multicolumn{2}{c}{(7)}
& (8) & (9) & (10) \\
\hline
 1 &11:40:36.44 &-26:24:09.9 &6.7 & 56.4&7.2& 5.0 & \multicolumn{2}{c}{}      &     &340&$<$-0.9\\
 2 &11:40:38.84 &-26:29:10.3 &3.6 & 9.2 & 3 & 0.8 & \multicolumn{2}{c}{}      &     &2.2&$<$-0.5\\
 3 &11:40:39.70 &-26:28:45.0 &3.5 &42.9&6.5 & 3.8 & \multicolumn{2}{c}{}      &     &6.0 & -0.8\\
 4 &11:40:44.21 &-26:31:29.8 &3.2 &11.2&3.3 & 1.0 &9.3 &3  & 5.4 &2.0 & -0.1\\
 5 &11:40:44.47 &-26:29:20.7 &2.4 & 12 &3.4 & 1.1 & \multicolumn{2}{c}{}      &     &4.6 & -0.4\\
 6 &11:40:45.96 &-26:29:16.7 &2.0 & 70 &8.4 & 6.2 & 11 &3.3& 6.4 &6.0 & -0.7\\
 7 &11:40:48.34 &-26:29:08.7 &1.5 &460 &21.5 &40.5&192 &14 &111.7&7.5 & -0.4\\
 8 &11:40:49.55 &-26:25:41.5 &3.9 &50.4&  7 & 4.4 &  \multicolumn{2}{c}{}  &     &51.6& -0.8\\
 9 &11:40:52.84 &-26:29:11.2 &0.6 & 58 &7.6 & 5.1 &  9 &3  & 5.2 &6.0 & -0.7\\
10 &11:40:54.21 &-26:29:43.5 &0.4 & 24 &5   & 2.1 &   \multicolumn{2}{c}{}  &     &7.5 & -0.7\\
11 &11:40:54.66 &-26:29:28.1 &0.2 & 67 &8.2 & 5.9 &   \multicolumn{2}{c}{}   &     &5.3 & -0.9\\
12 &11:40:58.17 &-26:30:27.1 &1.2 & 15 &3.9 & 1.3 &   \multicolumn{2}{c}{}   &     &4.4 & -0.3\\
13 &11:40:59.24 &-26:32:03.2 &2.8 &31.2&5.6 & 2.6 &19.0&4.3&10.0 &26.4& -0.2\\
14 &11:40:59.45 &-26:31:56.2 &2.7 &129.5&11.3&11.3&28.5&5.3&16.5 &31.0& -0.6\\
15 &11:40:59.70 &-26:32:07.5 &2.5 & 11 &3.3 & 1.0 &   \multicolumn{2}{c}{}     &     &9.2 &$<$0.6\\
16 &11:41:02.43 &-26:27:45.0 &2.3 &19.5&4.4 & 1.7 &13.0&3.6& 7.6 &85.0& -0.2\\
17 &11:41:03.00 &-26:27:34.2 &2.5 &22.6&4.8 & 2.0 &  \multicolumn{2}{c}{}      &     &107&$<$-0.8\\
18 &11:41:03.93 &-26:30:48.5 &2.4 & 64 &8   & 5.6 & 126&11.2&73.3&24.5& +0.3\\ 
\hline \hline
\end{tabular}
\end{center}
\footnotesize \noindent Notes: (1) Number in X-ray point source
catalog (2\&3) Right Ascension and declination in J2000 coordinates
(4) Distance in arcminutes from aim point of telescope (5) Net photon
counts in the soft band (0.5--2 keV) (6) X-ray flux in 10$^{-15}$
\ecs\ in the soft band assuming a power law spectrum with photon index
$\Gamma = 2.0$ (7) Net photon counts in the hard band (2--10 keV) (8)
X-ray flux in 10$^{-15}$ \ecs\ in the hard band assuming a power law
spectrum with photon index $\Gamma = 1.8$ (9) Aperture in arcsec$^2$
used for the photometry (10) Hardness ratio (H$-$S)/(H+S)
\end{table*}

\subsection{Photometry and astrometry} 

For each detected source with S/N $> 3$ in the soft band we report in
Table \ref{xray}: the source coordinates, the aim point distance in
arcminutes, the net photon count for the 0.5--2 keV (soft band S) and
the 2--10 keV (hard band H) energy ranges, the corresponding flux, the
aperture in arcsec$^2$, and the hardness ratio (H$-$S)/(H+S).

The photometry was performed through elliptical apertures, with
properties (i.e.\ sigma and PA) given by the region file produced by
WAVDETECT\footnote{These apertures are not the same as the WAVDETECT
source cells, which have arbitrary shapes.  However the photometry
output of WAVDETECT through these cells in the soft band agrees 
very well with the manual photometry through the elliptical apertures.  
This indicates that, because our sources are almost
all quite compact, the shape of WAVDETECT apertures can be
well approximated by ellipses.}.  The minimum size of the apertures is
$\sim$ 1\farcs6, implying that at energies below 6 keV, which includes
the majority of the detected photons, more than 80\% of the energy of
the X-ray sources is included (see Fig.\ 4.15 in Dobrzycki \etal\
2001).

The counts reported in Table \ref{xray} are background subtracted and
corrected for vignetting.  The background counts are negligible for
most sources.  We estimate the background in the soft and hard band to
increase the counts by 0.01 and 0.04 per pixel, respectively. In the
soft band this amounts to one background count for a 100 pixel
aperture, implying that for all detected sources (except \#1), the
number of background counts expected are less than one.  In the hard
band the background counts are not negligible for the largest sources.
The vignetting correction was done for the sources located $>
2$\arcmin\ from the aim point of the telescope. Fig.\ 4.4 of the {\it
Chandra} Proposer's Observatory Guide shows that the telescope
effective area decreases almost linearly with off-axis angle, averaged
over azimuth.  Following Cappi \etal\ (2001), we approximate the
curves of effective area for photon energies of 1.5 keV and 4.5 keV
with linear functions and use these to correct the soft and hard band
counts, respectively.  Note that the corrections are $< 4$\% for all
sources except for \#1.

The net count rates were converted into fluxes assuming a power law
spectrum with a Galactic absorption column along the line of sight of
$N_{\rm H} = 4.5 \times 10^{20}$ cm$^{-2}$ (Stark \etal\ 1992) and a
standard photon index of $\Gamma = 2.0$ for the soft band and $\Gamma
= 1.7$ for the hard band.  These spectral models were chosen to allow
a comparison with other observations such as the {\it ROSAT} and {\it
Chandra} Deep Fields. The total fluxes are only weakly dependent on
the spectral slopes and Galactic column density assumed.

Astrometric positions were initially determined from the ACIS-S aspect
solution. After the identification of the X-ray sources with objects
on a deep, ground-based $I$ band image discussed in Sect.\ 3, we refined
the relative astrometry between the X-ray frame and the optical image.
We used the optical and X-ray positions of the 6 sources which are
unresolved in both images and not saturated on the $I$ band: these
include the nucleus of MRC 1138-262 and two more sources with secure
redshift identifications (see below). Subsequently, we registered the
images using a general geometric transformation, resulting in a
relative astrometric accuracy of approximately 0\farcs1.  The results
are consistent with the shift of 0\farcs5$\pm$0\farcs1 in right
ascension and -0\farcs2$\pm$0\farcs1 in declination that Carilli
\etal\ (2002) applied to the image based on the position of the
nucleus of 1138-262 alone (e.g.\ for source \#7 which is the radio
galaxy nucleus, the shifts are 0\farcs45 and -0\farcs14 respectively).
The absolute astrometry of the $I$ band image was 
calibrated using 18 stars of the USNO catalog
(Monet \etal\ 1998) resulting in an
accuracy better than 0\farcs2.

\section{Optical identification of serendipitous X-ray sources}

Deep multicolour optical and near infrared images of the field were
obtained with the VLT FORS and ISAAC instruments during 2000 and 2001.
Details of the observations are reported in Kurk \etal\ (2002a).  To
identify the X-ray sources we have used the $I$ band image, which is the
deepest available having a limiting magnitude of 26.5 (except at the
edges of the field) and was obtained with a seeing of 0.65$''$. 
The total field of view of the dithered FORS
image is 0.016 deg$^2$ and its intersection with the ACIS S3 chip is
0.0154 deg$^2$; the NIR information is only available for the central
part of the field.

One of the X-ray point sources (\#1) lies outside the region covered
by our FORS image, amongst others due to a difference in the
orientation of the observations.  All other sources are identified in
the $I$ band image, with the exception of two that lie in the arms of a
nearby spiral galaxy.  The coordinates of all optical counterparts
deviate less than $\sim$ 1\farcs2 from the X-ray coordinates and most
of the sources are coincident within the astrometric uncertainty.
Given the depth of our $I$ band observations, the probability that one
of the X-ray sources coincides with an optical source not associated
with the X-ray source is non-negligible, about 9\%, even using a small
error circle of 1\farcs2 radius, corresponding to the largest
separation between the X-ray sources and their optical
identifications.  This means that one (1.35) of the 15 identifications
could be spurious.  However, all but one of the optical counterparts
are substantially brighter (2--3 magnitudes) than the detection limit
of the VLT image. At these magnitudes, the probability that an X-ray
source coincides with an unassociated optical object is much smaller
and their identification is thus more secure.

\begin{table*}
\begin{center}
\caption{Optical identifications of X-ray sources} \label{xray_opt}
\begin{tabular}{crrcccccccc} \hline \hline
N &  $\Delta$ RA  & $\Delta$ Dec       &m$_{\rm I}$ & $B-I$ & S  & Redshift &  L$_x$ & $\log({S_{x} \over S_{opt}})$ & ID\\
(1)  & (2)        & (3)                 & (4)        & (5) &(6) & (7)      &  (8)   & (9)                          &(10) \\
\hline
1  &      &        &&\multicolumn{3}{l}{Outside I-band image} &     & \\
2  &-0.12 & -0.07  & 18.6 &  5.2 & 0.99 &           &                & -1.3 &    \\
3  & 0.12 & -0.06  & 22.9 &  1.4 & 0.98 & 2.185     &  1.5           &  0.5 & AGN, Ly$\alpha$ emitter \\
4  &-0.70 &  0.23  & 23.1 &  3.9 & 0.02 &           &                &  0.5 & \\
5  & 0.09 &  0.01  & 25.3 &  1.6 & 0.52 & $\sim$ 2  &  0.43          &  0.7 & ERO ($I-K >$ 5.3)  \\
6  & 0.01 & -0.07  & 20.8 &  1.8 & 0.98 & 2.157    &  2.5           & -0.2 & AGN, H$\alpha$ emitter  \\
7  & 0.00 &  0.06  & 22.0 &  1.3 & 0.05 & 2.156     &  16.3          &  1.2 & Radio galaxy  \\
8 & 0.88 &  0.10  & 16.6 &  3.4 & 1.00 & 0         &                & -2.4 & star? \\
9 &-0.10 &  0.04  & 21.7 &  1.6 & 0.98 & $\sim$2.16&  2.1           &  0.0 &\lya\ emitter& \\
10 & 0.10 & -0.03  & 17.8 &  2.7 & 0.03 &           &                & -1.8 &spiral galaxy \\
11 &-0.03 & -0.0   & 17.1 &  4.4 & 0.99 &           &                & -1.4 &  \\
12 &-0.34 & -0.11  & 20.3 &  2.8 & 0.03 &           &                & -0.9 &elliptical \\
13 &      &        & &\multicolumn{3}{l}{in spiral$^a$} \\
14 &-1.03$^{b}$ & -0.07$^{b}$  & 15.9 &  2.5 & 0.02 &           &    & -1.5 &IRAS/spiral gal.\\
15 &      &        &  & \multicolumn{3}{l}{in spiral$^a$} \\ 
16 & 0.51 & -0.19  & 22.9 &  1.7 & 0.19 & $\sim$2.16&   0.67         &  0.1 & \lya\ emitter   \\
17 & 0.58 &  1.03  & 18.5 &  4.9 & 0.99 &           &                & -1.1 &   \\   
18 &-0.74 &  0.02  & 21.1 &  2.9 & 0.16 &           &                & -0.1 &merger$^{c}$ \\\hline \hline
\\
\end{tabular}
\end{center}
\footnotesize \noindent Notes: (1) Number in X-ray point source
catalog (2\&3) Distance between X-ray source and optical
identification in RA and Dec (both in arcseconds) (4) $I$ band magnitude
(5) B$-$I colour (6) $I$ band stellaricity index from Sextractor: from
0.00 (extended source) to 1.00 (point source) (7) Redshift (either
secure or an estimate) (8) X-ray luminosity in 10$^{44}$ erg s$^{-1}$
(9) Logarithm of the X-ray to optical flux ratio (10) Optical
identification 
\\$^a$ There are three sources inside the optical extent of one nearby
spiral galaxy; \#14 coincides with the nucleus
\\$^b$ The nucleus of this bright spiral is saturated on the optical
image, so the optical position is known only with some uncertainty
\\ $^c$ Merger or interacting system with tidal arms
\end{table*}

The optical identifications are listed in Table \ref{xray_opt}, where we
report the difference in position between the X-ray source center and
the optical counterpart center (as determined by Sextractor, Bertin \&
Arnouts 1996), the total $I$ band magnitude, the $I$ band stellaricity
index from Sextractor, which indicates the extendedness of the object
(1.0 is unresolved, 0.0 is extended), the redshift if available or an
estimate of the redshift, the implied X-ray luminosity, and the soft
X-ray to visual band flux ratio computed using $R$ band magnitudes.  We
cross-correlated the X-ray sources with our catalog of objects with
excess emission in narrow band filters corresponding to \ha\ and/or
\lya\ at $z \sim 2.16$, i.e.\ the redshift of the radio galaxy (Kurk
\etal\ 2000) and we report any such identification in Table \ref{xray_opt}. We
have also checked our deep 1.4 GHz VLA radio images of this field
(unpublished) to search for possible radio counterparts: besides the
central radio galaxy we detect weak unresolved emission only from the
nucleus of the spiral galaxy (object \#14).

Here we briefly discuss some characteristics of individual sources.
\\
{\it Source \#3} is identified with a faint QSO (\lya\ emitter
\#1687 in Pentericci \etal\ 2000) at $z \sim 2.185\pm0.005$ (where the
uncertainty is mainly due to the presence of absorption features),
showing broad \lya, \ion{Si}{iv} and \ion{C}{iv} emission lines in its
optical spectrum.
\\
{\it Source \#5} is identified with a faint extended object, having a
color $R - K = 5.8$ ($I - K = 5.3$) such that it can be classified as
an Extremely Red Object (ERO, Elston \etal\ 1988).  A spectrum of
this object has not yet been obtained,
but its colors indicate that it could be a 
galaxy containing an evolved stellar population at
$z > 1.5$ or a dust reddened starburst galaxy at $z \sim 2$ (see the
discussion in Kurk \etal\ 2002a).
\\
{\it Source \#6} is identified with another faint quasar at $z =
2.157\pm0.002$ showing very broad \ha\ emission in the near infrared
spectrum (Kurk \etal\ 2002b).
\\
{\it Source \#7} is the nucleus of the radio galaxy MRC 1138-262 at
$z = 2.156$ and is extensively discussed in Carilli \etal\ (2002).
\\
{\it Source \#9} is identified with a faint point source which also has
excess \lya\ emission in the narrow band images corresponding to
redshift $\sim$ 2.16.  No spectrum has been obtained yet.
\\
{\it Source \#14} coincides with the nucleus of a spiral galaxy, most
probably identified with IRAS F11384--2615 which has been detected by
IRAS at both 60 and 100 $\mu$m (Moshir \etal\ 1990).
\\
{\it Source \#16} is a candidate \lya\ emitter with very strong and
extended \lya\ emission. It is very likely another AGN at $z = 2.16$
encompassed by a \lya\ halo (Kurk \etal\ 2002a). Because this object is at the very edge
of the $I$ band image, it was not included in the original sample of
candidate \lya\ emitters (Kurk \etal\ 2000) and an optical spectrum
was not obtained.

Of the other sources, \#10 is associated with the nucleus of a
spiral galaxy; \#13 and \#15 lie in the arms of the spiral galaxy \#14
and cannot be identified in the optical bands.  Four sources (\#2,
\#8, \#11 and \#17) are identified with unresolved objects with $I$ band
magnitudes in the range 16.6--18.6. The brightest of these might be a
star as suggested by the X-ray to optical spectral index (see below).
Source \#12 coincides with a 20$^{\rm th}$ magnitude
elliptical galaxy and at the Eastern edge of our $I$ band image there is
a merging system consisting of two extended objects, one of which is
quite elongated and exhibits tidal arms.  This object coincides with
source \#18.

\begin{figure*}
\epsfig{file=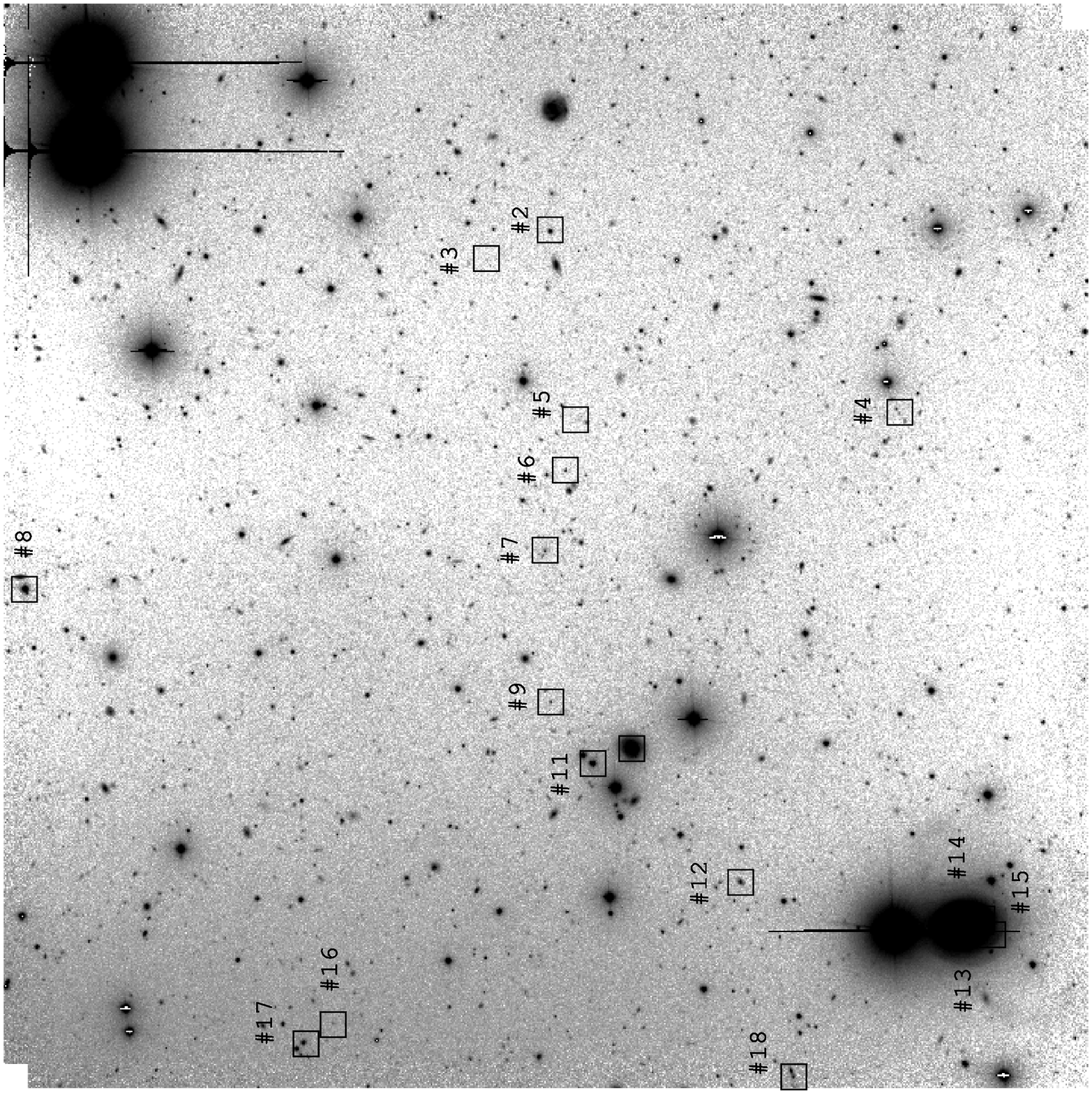,height=17.5cm,clip=,angle=-90} 
{\bf Figure 1} An $I$ band image of the FORS field (approximately 6.8$'
\times$6.8$'$) around the radio galaxy MRC 1138-262.  The X-ray
sources with optical identifications are indicated by squares and
labelled as in Table 1.  Source \#7 is the radio galaxy. Sources \#13
and \#15 are within the spiral arms of galaxy \#14.
\end{figure*}

\section{Analysis}
\subsection{The source counts}

In Table \ref{xray_counts}, we present the number of sources found in
the FORS field and the source densities per square degree with fluxes
$> 1, 1.5$ and $3 \times 10^{-15}$ erg cm$^{-2}$ s$^{-1}$ in the soft
band and the total number in the hard band.  We find respectively 16,
12, 8 and 4 sources.  It is instructive to compare our results with
those of other deep {\it Chandra} observations.  Therefore, we also
report the densities found in two recent {\it Chandra} Deep Fields
(Mushotzsky \etal\ 2000; Giacconi \etal\ 2001), and the number of
sources expected  in the FORS field from these statistics.  The table
shows that the field of MRC 1138-262 contains about 50\% more sources
than the CDF in each of the three flux bins in the soft band  
(with a significance of about 1.5$\sigma$ in each flux bin).

There are large field-to-field variations observed in surveys of the
resolved components of the X-ray background, due to fluctuations of
the large scale structure (the ``cosmic variance'').  Although the
excess in the field of MRC 1138-262 is only 50\%,
this is still larger than the 20--30\% observed amplitude of the cosmic
variance (e.g.\ Cappi \etal\ 2001). Moreover, no comparable excess is
observed in the hard band, although the statistics are poorer: the
density of the four detected objects with flux $> 10^{-14}$ erg cm$^{-2}$
s$^{-1}$ is $\sim 250$ sources deg$^{-2}$, consistent with the CDF
results of 200--300 sources deg$^{-2}$.

We conclude that we have found a true over-density of soft X-ray
sources in the field around MRC 1138-262.  The density is similar to
that revealed by Cappi \etal\ (2001) in two clusters of galaxies at $z
\sim 0.5$: 3C295 and RXJ0030.  They also find this surplus only in the
soft band and not in the hard band.  Their most likely explanation of
the over-density is that the clusters are atypical and contain more
AGN than other distant clusters. 
No other similar studies have been reported on the
highest redshift known clusters ($ z \sim$ 1--1.2) which have been
observed with {\it Chandra}. We will discuss this further in the
last section. 

\begin{table}
\begin{center}
\caption[]{X-ray point source counts} \label{xray_counts}
\begin{tabular}{c c r@{/}l r@{$\pm$}r r@{/}l}
\hline \hline
Flux & Total & \multicolumn{2}{c}{Exp CDF} & \multicolumn{2}{c}{n(1138)} & \multicolumn{2}{c}{n(CDF)} \\
(1) & (2) & \multicolumn{2}{c}{(3)} & \multicolumn{2}{c}{(4)} & \multicolumn{2}{c}{(5)}\\ \hline
 \multicolumn{4}{l}{Soft band (0.5--2 keV)} \\
1.0 & 16 & 10.3&11 & 1040&260 & 670&720  \\   
1.5 & 12 & 7.2&8.3 &  780&220 & 470&540  \\
3.0 &  8 & 4.0&5.1 &  520&180 & 260&330  \\ \hline
\multicolumn{4}{l}{Hard band (2--10 keV)} \\
10  &  4 & 3.7&5.4$^a$ & 260& 130  & 240&350$^a$ \\ \hline \hline
\end{tabular}
\end{center}
\footnotesize \noindent Notes: (1) Minimum source flux in 10$^{-15}$
\ecs\ (2) Number of X-ray sources detected in the FORS field of MRC
1138-262 (3) Number of X-ray sources expected in a FORS-sized field
from {\it Chandra} Deep Field counts; the two values are from the
South and North surveys respectively (Giacconi \etal\ 2001, Mushotzsky
\etal\ 2000) (4) Density with Poisson error (deg$^{-2}$) of X-ray
sources in the field of MRC 1138-262 (5) Density (deg$^{-2}$) of X-ray
sources from the two {\it Chandra} Deep Field surveys\\ $^a$ the CDF
North result was derived using $\Gamma = 1.4$ and not $\Gamma = 1.7$.
\end{table} 

\subsection{Properties of the serendipitous sources}
Clues on the nature of the field X-ray sources come from their
hardness ratios, optical morphologies and optical to X-ray spectral
indices.

Due to the paucity of photons in individual sources, their spectral
properties can best be studied by the hardness ratio (HR).  The
distribution of the hardness ratios as a function of the soft band
flux is shown in the lower panel of Fig.\ 2.  The HR of the sources
indicate that most are soft, as expected from objects initially
detected in the soft band, with the exception of source \#18.  The
average value of the HR is -0.51, similar to that observed for type I
AGN with a broad range of redshifts (e.g. Tozzi \etal\ 2001).  The
hardness ratio does not appear to depend on the soft X-ray flux,
consistent with Tozzi \etal\ (2001), who show that the number of hard
sources increases only at fluxes lower than our limit of 10$^{-15}$
erg s$^{-1}$ cm$^{-2}$.  The only hard source in the sample is \#18:
as noted in Sect.\ 3 it coincides with the faintest of a pair of
interacting objects. Its HR is consistent with a type II AGN at
intermediate redshift ($z \sim 1$) with intrinsic absorbing column of
the order of $N_{\rm H} \sim 10^{23}$ cm$^{-2}$ (assuming that it has
an average spectrum with $\Gamma = 1.7$).

\begin{figure}
\epsfig{file=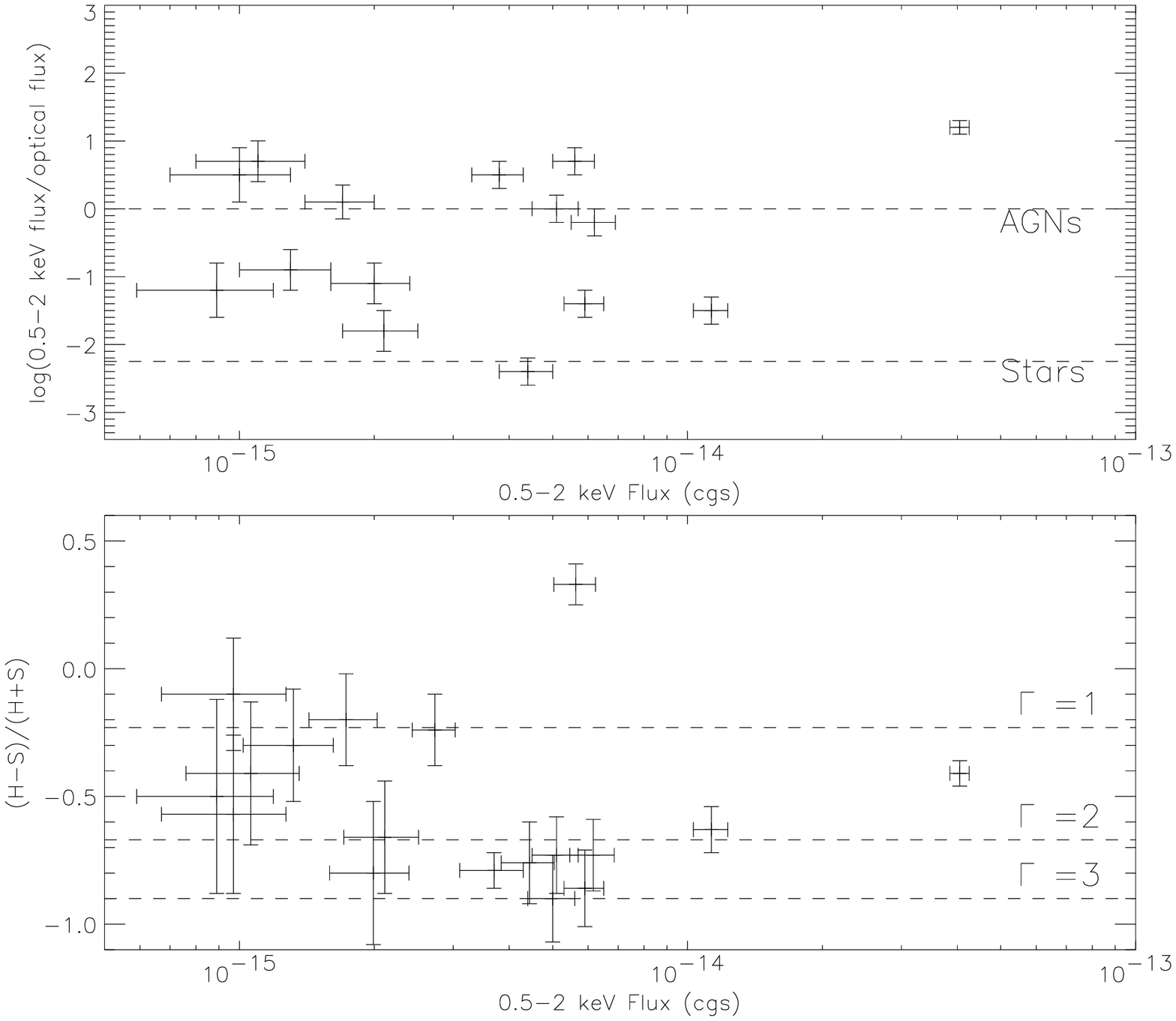,height=8.80cm,clip=,angle=90}
{\bf Figure 2} Upper panel: the soft X-ray to optical flux ratio
versus the soft X-ray flux. The dotted lines give the typical ratios
for AGN (upper) and stars (lower) as determined by Stocke \etal\
(1991).  Lower panel: the hardness ratio (H$-$S)/(H+S) versus the soft
band flux for all sources. The dashed lines show power law models with
different photon indices ($\Gamma$) computed assuming the galactic value
$N_{\rm H} = 4.5 \times 10^{20}$ cm$^{-2}$.
\end{figure}

How do the optical counterparts of the X-ray sources compare with
those found in the literature?  We find that approximately one third
of the sources are extended and two thirds of the sources appear
pointlike. These numbers are similar to those
found by Giacconi \etal\ (2001).  Note that the resolution of the
images is about 0\farcs6-0\farcs7, which at redshift 2 corresponds to
5--6 kpc, i.e.\ not a very stringent limit. Even starburst galaxies at
moderately high redshift are expected to be unresolved at this
resolution.

In the upper panel of Fig.\ 2, we plot the X-ray to $R$ band flux ratio
as a function of X-ray soft band flux: the values are in general
consistent with those found by Lehmann \etal\ (2001) in the {\it
ROSAT} Deep Survey, for sources with a similar range of soft band
fluxes.  They found that most of the AGN population (both broad line
and narrow line AGN) cluster around a best fit line S$_x$/S$_o = 1$
(or $\log (S_x/S_o) = 0$).  Similarly Stocke \etal\ (1991) find that
AGN all have a logarithmic ratio $> -1$, whereas objects with a
logaritmic ratio $< -2$ are probably galactic stars.  Our
plot indicates that most of the objects have properties consistent
with AGN, while object \#8 is most probably a star.  

Tozzi \etal\ (2001) find a population of soft (HR $\sim -1$), low
luminosity sources with low X-ray to optical ratio and X-ray soft band
flux $< 3 \times 10^{-15}$ erg s$^{-1}$ cm$^{-2}$ which appear to be
galaxies (as opposed to AGN) and are located at low or intermediate
redshifts with an average $\bar{z} \sim$ 0.2--0.3.  Their X-ray
emission is probably produced by low power nuclear activity.  In 
our sample, three sources (\#2, \#10 and \#17) have low enough fluxes
(they are detected only in the soft band) and X-ray to optical flux ratio  
($\log  (S_x/S_{o}) \le -1$)
to satisfy the above criteria (see upper panel of Fig.\ 2). Source
\#10 is indeed identified with a low redshift spiral galaxy, whereas
the other two are unresolved so they could be small intermediate
redshift galaxies hosting low power AGN.

\section{Discussion}

As noted in Sect.\ 4.1, there is an excess of soft X-ray
sources as compared to deep field counts.  The origin of this excess
can be explained by the number of X-ray sources which are or could be
associated with the $z \sim 2.16$ protocluster around the radio galaxy
MRC 1138-262 (Pentericci \etal\ 2000).

Indeed the redshifts of two of the X-ray emitters (\#3 and \#6) have
been confirmed spectroscopically.  For two more (\#9 and \#16), there
is a very good indication of their redshift from the presence of
excess flux in the optical narrow band corresponding to \lya\ emission
at the redshift of the radio galaxy.  Source \#5, the ERO, has colours
consistent with a redshift $\sim$ 2 object, so it could also be part of the
protocluster.  Finally, it is possible that other objects, in
particular the optically fainter sources such as \#4, are also located
in the protocluster. Plausibly then, 5 or 6 sources are part of the
protocluster at $z = 2.16$. Subtracting this number from the total
detected sources in the field of MRC 1138-262, the number of remaining
sources are fully consistent with the Deep Field measurements in all
flux bins.  None of the sources at $z \sim 2.16$ are detected in the
hard X-ray band, so the statistics for this band also remain 
consistent with the Deep Field results.
 
For the remainder of this discussion, we will assume that these
sources are indeed all at redshift $\sim$ 2.16.  What powers the X-ray
emission in the sources associated with the cluster? Are they AGN or
powerful starburst galaxies?  Based on the X-ray and optical
properties discussed in the previous section, we conclude that they
are most probably all AGN.  Furthermore, the implied X-ray
luminosities are far too large to be due to starbursts. The optical
and NIR spectra of the two sources with known redshift, showing very
broad emission lines (with widths of several 1000 km s$^{-1}$), also
confirm that they are AGN.

Is it unusual to have such a high concentration of AGN at redshift
$\sim$ 2?  We know that the density of AGN is a strong function of
cosmic epoch: at $z = 2$ they were a hundred times more common than in
the present universe (e.g.\ Boyle \etal\ 2000), so perhaps it is not
surprising to find several of them in a $z = 2.16$ protocluster.  From
the soft X-ray luminosity function of AGN we can estimate how many
sources we expect in a given region.  At redshift $\sim$ 2 the number of sources 
with luminosities $\log L_x$ between 43.6 and 45 is $6.4 \times 10^{-4}$
Mpc$^{-3}$ (according to Table 5 in Miyaji \etal\ 2001).  For the 4758
Mpc$^3$ volume corresponding to the FWHM of the filter used for the detection of
Ly$\alpha$ emitters (Kurk \etal\ 2000), three AGN would be
expected. The protocluster around MRC 1138-262 contains about twice
the number expected, which is quite unusual since normally clusters
contain {\sl less} AGN than the field.  For example, in the nearby
universe AGN occur rarely in clusters, comprising only 1\% of all
cluster galaxies (regardless of morphology, see Dressler \etal\ 1985),
while they are much more common in the field population (about 5\%,
e.g.\ Huchra \& Burg 1992).  There is no evidence for an increase with
redshift of the AGN fraction in clusters up to $z \sim 0.5$ (Dressler
\etal\ 1999).  We must point out, however, that most studies of this
kind have been conducted with optical surveys where the dusty AGN or those
with weak emission lines may not be recognized as such. 
Indeed, recent observations in other parts of the
spectrum, including radio and X-rays, have shown unusual abundances of
AGN in several clusters: Molnar \etal\ (2002) recently reported an 
excess number of X-ray sources around the cluster A1995 (z$=$0.32), and  
we already mentioned 3C295 at $z = 0.46$ and
RX0030 at $z = 0.5$ (Cappi \etal\ 2001; Dressler and Gunn 1983). In
addition, Best \etal\ (2002) found a higher proportion of radio
sources identified with AGN in the rich cluster MS1054-03 at $z =
0.83$ than in the field.  Finally, {\it Chandra} observations of the
cluster Abell 2104 at $z = 0.154$ reveal an unexpectedly high AGN
fraction (Martini \etal\ 2002).  Five out of six AGN in this cluster
would not have been classified as such based on their optical spectra.
In conclusion, we do not yet have proper statistics on the AGN fraction 
in clusters and on how this evolves up to the redshift 
of MRC 1138-262.

Finally, we draw attention to the string of X-ray emitters formed by
the radio galaxy (\#7) and sources \#3, \#5, \#6 and \#9, roughly
along the East--West direction, as shown in Fig.\ 1.  This string of
X-ray sources, containing objects at redshifts close to that of the
radio galaxy, is aligned with the axis of the radio emission of the
central galaxy and with the general distribution of \ha\ emitters in
the cluster (see Kurk \etal\ 2002a).  The extended X-ray emission
associated with MRC 1138-262, which we interpreted as produced by gas
shock heated by the expanding radio source, is also aligned along
this axis (Carilli \etal\ 2002).  All observations suggest that the
ambient medium (gas and galaxies) is anisotropically distributed
and that the radio jet has emerged along this direction, or is
preferentially detected if it propagates along the direction of
highest ambient density.  As it seems, the ambient medium is
distributed similarly both on small scales (100 kpc) and large scales
(1--2 Mpc).  West (1994) has extensively discussed a hierarchical
formation scenario of massive cD galaxies being built up from smaller
protogalactic units, resulting in a remarkable coherence of structures
from the central engines of AGN to the large scale structure of the
universe.

\section{Conclusions}

We have reported on the serendipitous X-ray emitters detected by {\it
Chandra} in the field around the radio galaxy MRC 1138-262. We have
presented optical identifications for 15 of the 18 sources from deep
VLT observations.  Based on the X-ray and optical properties of the
sources, we conclude that most are type I AGN, one is a
type II AGN and one is probably a star.  Compared to the $\log N -
\log S$ relation measured in deep {\it Chandra} observations, we find
an excess of soft X-ray sources in the field of MRC
1138-262.  The most probable explanation is that several of the X-ray emitters
are associated with the $z \sim 2.16$ protocluster around the radio
galaxy. Indeed, two of these have been confirmed spectroscopically to
be broad emission line AGN at a redshift similar to that of the
central galaxy.  Additional spectroscopy is needed to determine how
many of the X-ray emitters are really associated with the $z \sim 2.16$
protocluster.  Furthermore, future observations of AGN in other
distant cluster of galaxies will help to clarify whether the
concentration of AGN around MRC 1138-262 is unusual or just typical
for such a structure at high redshift.

\begin{acknowledgements}
The National Radio Astronomy Observatory (NRAO) is operated
by Associated Universities, Inc. under a cooperative agreement with the
National Science Foundation.
CC and DH acknowledge support from grants from the Chandra
X-ray Center. Work at SAO was partially supported by NASA grant GO0-1137B and contract
NAS8-39073.     
\end{acknowledgements}

\end{document}